\documentclass[apj]{emulateapj}

\shorttitle{3D variable GRB simulations}
 
\shortauthors{Lopez-Camara et al.}

\begin{document}

\title[Time-scales in long GRB simulations ]{Three-dimensional simulations of long duration gamma-ray burst jets: time scales from variable engines}

\author{D. L\'opez-C\'amara\altaffilmark{1, *}, Davide Lazzati\altaffilmark{2}, Brian J.  Morsony\altaffilmark{3}}

\altaffiltext{1}{Instituto de Astronomía, Universidad Nacional
  Autónoma de México, Apdo. Postal 70-264, Cd. Universitaria, México
  DF 04510, México}

\altaffiltext{2}{Department of Physics, Oregon State University, 301
  Weniger Hall, Corvallis, OR 97331, USA}

\altaffiltext{3}{Department of Astronomy, University of Maryland, 4296
  Stadium Drive, College Park, MD 20742-2421, USA}
  
\altaffiltext{*}{diego@astro.unam.mx}

\begin{abstract}
Gamma-ray burst light curves are characterized by marked variability, each showing unique properties. The origin of such variability, at least for a fraction of long GRBs, may be the result of an unsteady central engine. It is thus important to study the effects that an episodic central engine has on the jet propagation and, eventually, on the prompt emission within the collapsar scenario. Thus, in this study we follow the interaction of pulsed outflows with their progenitor stars with hydrodynamic numerical simulations in both two and three dimensions. We show that the propagation of unsteady jets is affected by the interaction with the progenitor material well after the break-out time, especially for jets with long quiescent times, comparable to or larger than a second. We also show that this interaction can lead to an asymmetric behavior in which pulse durations and quiescent periods are systematically different. After the pulsed jets drill through the progenitor and the interstellar medium, we find that, on average, the quiescent epochs last longer than the pulses (even in simulations with symmetrical active and quiescent engine times). This could explain the asymmetry detected in the light curves of long-quiescent times duration gamma-ray bursts.

\end{abstract}


\section{Introduction}\label{sec:intro}
All the observed gamma-ray burst (GRB) light curves (LCs) are characterized by temporal variability \citep{walker00}, and some GRBs present pulses separated by quiescent time intervals lasting from fractions of a second to several tens of seconds \citep{fenimore00,borgonovo07}. Most, if not all the variability is ascribed to the fact that the relativistic outflow is injected by an unsteady central engine that produces outflows with varying rest-frame density, Lorentz factor, luminosity, magnetization, or of a combination of them \citep{p90, rm94, rr01a, rr01b}. Additional variability can be introduced by the interaction of the outflow with the progenitor star \citep{mor10}, by turbulence \citep{nk09}, or by magnetic reconnection (e.g. the ICMART model, \citet{zy11, zz14}).

\citet{nakar02} (NP02) analyzed the GRB dataset from the BATSE 4B catalogue, specifically the temporal behavior of the 68 brightest long GRBs. The total number of pulses and quiescent times in these bright bursts was in total 1330 and 1262 (respectively). NP02 find that the pulses follow a lognormal distribution while the intervals do not and conclude that the pulses and quiescent epochs come from different mechanisms. Whether the pulses and quiescent periods can be due to the same mechanism or not remains to be fully studied. In particular, NP02 did not take into account the fact that the jet propagation through the progenitor star can modify the temporal variability injected by the engine (e.g., \citet{mor10, laz11}). To shed more light on this phenomenon, we performed a series of two- and three-dimensional simulations in which a central engine injects an unsteady outflow at the center of a massive progenitor star. We follow the jet as it propagates through the progenitor star and the interstellar medium (ISM) and eventually calculate LCs and study the behavior of pulses and intervals in between.

This paper is organized as follows. We first describe the simulation setup, and the numerical models in Section~\ref{sec:input}, followed by discussion of our results in Section~\ref{sec:results}. Conclusions are given in Section~\ref{sec:conc}.

\section{Simulations setup and numerical models}\label{sec:input}
We performed a series of three-dimensional (3D) and two-dimensional (2D) simulations, each consisting of a variable GRB jet drilling through the stellar progenitors envelope and then through the interstellar medium (ISM). All simulations were performed using the hydrodynamic code FLASH (version 2.5) in Cartesian (3D) or cylindrical (2D) coordinates \citep{fryx00}. The setup was analogous to the one in \citet{lc13}, but the jets studied in this work are intermittent, produced by engines with a diverse set of duty cycles. Given the computational demand in order to follow the jet propagation in 3D and the need of exploring a big parameter space, several two-dimensional models were also run, this after verifying their reliability against the 3D results.

In 3D, jets from four intermittent engines with different active/quiescent periods were modeled. The engine period was constant through the simulation and the engines were characterized by square pulses with active and quiescent durations equal to each other. The numerical setup was very similar to that from \citet{lc13}. The variable jet had average luminosity $\langle L\rangle=5\times10^{50}$~erg~s$^{-1}$, opening angle $\theta_0=10^{\rm{o}}$, initial Lorentz factor $\Gamma_0$=5, ratio of internal over rest-mass energy $\eta_0$=80 (with which $\Gamma_{\infty}$=400), and was launched from the core of the progenitor (at r$_{\rm{in}}$=10$^{9}$~cm). The stellar progenitor, model 16TI from \citet{wh06}, was immersed in a constant density ISM ($\rho_{\rm{ism}}$=10$^{-10}$~g~cm$^{-3}$). Engines with period $T=$0.2, 1, 2, and 4~s were simulated. Real GRB engines are likely characterized by faster variability, but we could not reduce the period without making the time step prohibitively small for the full 3D simulations. A control simulation was also run with a jet that was always active, and in order to check for resolution effects an extra simulation, akin to the 3D variable jet model with a 0.1s period, was performed but with the entire cocoon been solved with the finest resolution grid level (indicated as ``HR'').

The numerical domain of the 3D simulations covered the top half of the pre-SN progenitor star as well as the ISM it is immersed in. The boundaries were set at 2.56$\times$10$^{11}$~cm along the polar axis (Y-axis in the figures) and $\pm$6.4$\times$10$^{10}$~cm in the equatorial plane (XZ-axis). Eleven levels of refinement were used with which the finest grid level resolution was $\Delta$=7.81$\times$10$^{6}$~cm, which is comparable or even an order of magnitude finer than previous 3D collapsar studies \citep{zwh04, w08, lc13, brom16, ito16}, and comparable with two-dimensional GRB-jet studies \citep{zwm03, zwh04, mor07, mor10, laz09, nag11, laz12, miz13}. The base of the jet (Y$<$0.5~r$_{\rm{in}}$) was solved with the finest resolution grid level, the lower and mid part of the jet (Y$<$4~r$_{\rm{in}}$) was followed with a resolution of 2$\Delta$, and the rest of the jet was followed with 4$\Delta$. As pointed out in \citet{lc14}, when launching a variable jet from the core of a pre-SN like progenitor the cocoon must be followed with resolutions comparable to that from the jet. Thus, the cocoon was followed with a resolution which covered values as fine as 4$\Delta$ (close to the polar axis) up to 32$\Delta$ (far from the polar axis). In the 3D HR model all of the cocoon was followed with 4$\Delta$.

3D simulations are very important in order to properly take into account the jet propagation inside a massive progenitor \citep{zwh04, lc13, brom16, ito16}, none the less they are very time-consuming. Thus, to be able to explore a wider parameter range and perform a more meaningful comparison of our results with observations, we performed a large set of 2D numerical simulations. The first set of 2D simulations replicated the 3D work, with constant square-pulsed intermittent engines characterized by periods $T=0.2, 1, 2,$ and $4$~s for a total integration time of 50~s. This set was intended to compare the results of our pulsed jet models between two- and three-dimensions. The second set of 2D simulations (twenty in total) had random active and quiescent interval values and was intended to provide a more realistic comparison with the observations. A summary with the main characteristics of the models is shown in Table 1. The 2D simulations were solved with the same resolution and numerical domain as the 3D LR models.

\begin{table}
\tabletypesize{\footnotesize}
\caption{Model characteristics}
\begin{center}
\begin{tabular}{cccccc}
  \hline
  Model & 3D/2D & T(s) & t$_{\rm{max}}$(s) & $\Delta$ & t$_{\rm{bo}}$(s) \\
  \hline
  m3D0.2lr & 3D & 0.2 & 17.40 & LR & 11.00 \\
  m3D1.0lr & 3D & 1.0 & 15.20 & LR & 7.80 \\
  m3D2.0lr & 3D & 2.0 & 17.13 & LR & 6.73 \\
  m3D4.0lr & 3D & 4.0 & 13.33 & LR & 6.80 \\
  m3D0.2hr & 3D & 0.2 & 12.80 & HR & 11.20 \\
  m3Donlr & 3D & always on & 7.80 & LR & 5.27 \\
  m2D0.2lr & 2D & 0.2 & 50.00 & LR & 5.60 \\
  m2D1.0lr & 2D & 1.0 & 50.00 & LR & 10.07 \\
  m2D2.0lr & 2D & 2.0 & 50.00 & LR & 11.93 \\
  m2D4.0lr & 2D & 4.0 & 50.00 & LR & 12.34 \\
  m2Dranlr$^*$ & 2D & random & 50.00 & LR & -  \\
  \hline
  $^*$Note: 20 models \\
  \hline
\end{tabular}
\end{center}
\label{default}
\end{table}

\section{Results and Discussion}\label{sec:results}

\subsection{Three-dimensional models}\label{sec:3D}
\begin{figure}
  \centering
  \includegraphics[width=0.9\linewidth]{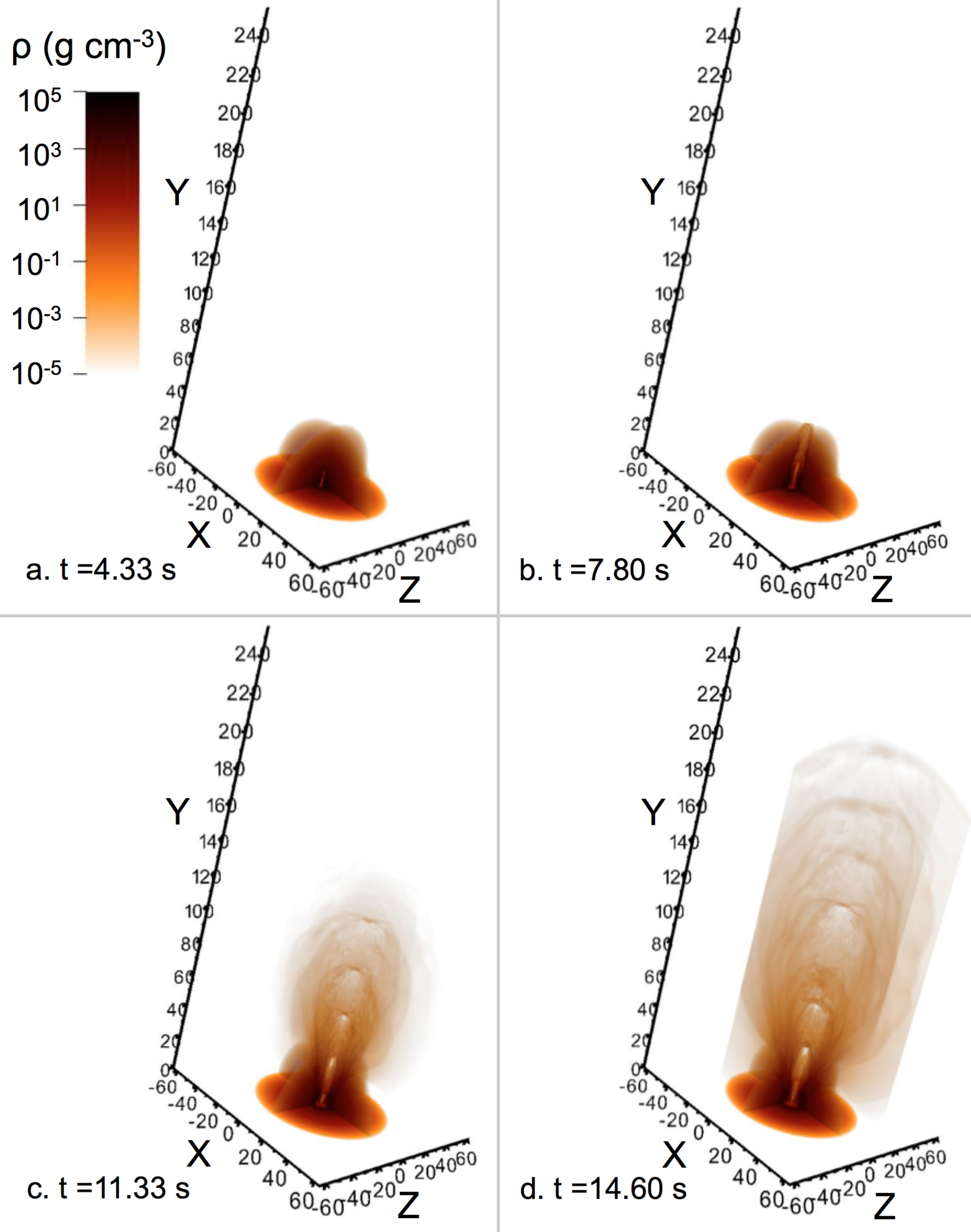}
  \caption{Density stratification maps (g~cm$^{-3}$) of the XY, ZY,
    and XZ planes for model m3D1.0lr ($T=1$~s) at different time
    frames (a. t=4.33s, b. 7.80s, c. 11.33s, and d. 14.60s). The axis are scaled
    to 10$^{9}$~cm.  In order to better visualize the density jumps, the minimum value in all the density stratification plots was set to 10$^{-5}$~g~cm$^{-3}$. (A color version of this figure is available in the online journal.)}
  \label{fig:fig1}
\end{figure}

All jets in the the 3D models successfully drilled through the stellar progenitor, broke out of the star, and evolved through the ISM, eventually leaving the numerical domain. The evolution of one of the pulsed models through the progenitor and then through the ISM is shown in Figure~\ref{fig:fig1}. Specifically, we show the density stratification maps (XY, XZ, and YZ planes) for the $T=1$~s model as it drills through the stellar envelope. It breaks out of the progenitor after 7.8~s, and at 14.60~s is about to reach the upper polar axis boundary.

\begin{figure}
  \centering
  \includegraphics[width=0.75\linewidth]{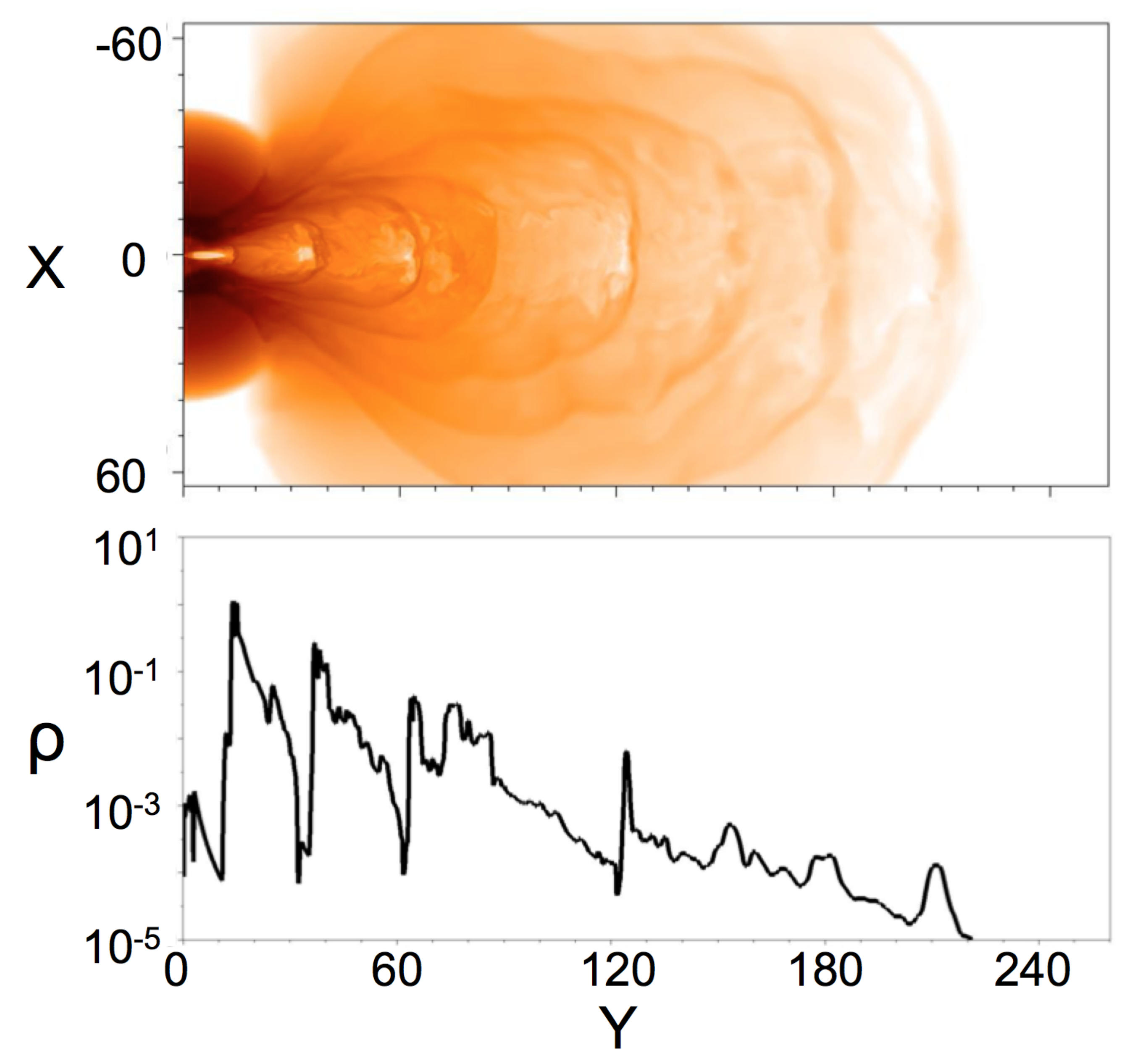}
  \caption{Density stratification map of the XY plane for model m3D1.0lr at 14.60s (axis and stratification map are scaled as in Figure~\ref{fig:fig1}) and the radial profile at the polar axis (upper and lower panel respectively). (A color version of this figure is available in the online journal.)}
  \label{fig:fig2}
\end{figure}

The pulsed behavior in all the models is distinguishable as density jumps in the jet. In order to further show such density jumps, in Figure~\ref{fig:fig2} we show the density stratification map in the XY plane and the radial density profile along the polar axis for the $T=1$~s period model. The density jumps are present inside the progenitor ($\sim$1$\times$10$^{10}$~cm), in the edge of the pre-SN progenitor ($\sim$4$\times$10$^{10}$~cm), and outside the progenitor ($\sim$6$\times$10$^{10}$~cm, $\sim$1.2$\times$10$^{11}$~cm, $\sim$1.8$\times$10$^{11}$~cm, and $\sim$2.1$\times$10$^{11}$~cm). The active pulsed epochs of the central engine produces pulses with low-density regions (with minimum values as low as $\sim$10$^{-4}$-10$^{-5}$~g~cm$^{-3}$) which drill through the pre-SN progenitor and propagate over the ISM. The pulses are separated from each other by quiescent periods with density values that are at least an order of magnitude higher ($\sim$10$^{0}$-10$^{-4}$~g~cm$^{-3}$) than the subsequent pulse.

The 3D pulsed models require more time in order to break out of the progenitor compared to the model where the jet was always active, models with a break out time equal to $t_{\rm{bo}}$=5.27~s. The 3D models with $T=0.2, 1, 2$, and 4~s have break out times equal to t$_{\rm{bo}}$=11.00s, 7.80s, 6.73s, and 6.80 respectively. These break-out times are consistent with what has previously been observed in recent 3D GRB jet studies \citep{lc13, brom16, ito16}, with 2D GRB jet studies \citep{zwm03, zwh04, mor07, laz09, mor10, nag11, laz12, miz13}, and with the analytic solution of \citet{brom11}.
Interestingly, the models with longer episodic periods tend to have shorter break out times and viceversa (t$_{\rm{bo}} \propto$T$^{-0.2}$). We notice that the break-out time of the $T=$0.2s simulations is t$_{\rm{bo}} \sim$11~s, independently of the resolution. This anti-correlation between the period of the engine and the breakout time is consistent with results from \citet{mor10} in 2D simulations (a more thorough comparison between 2D and 3D breakout times is treated in Section~\ref{sec:2D}).

\begin{figure}
  \centering
  \includegraphics[width=0.8\linewidth]{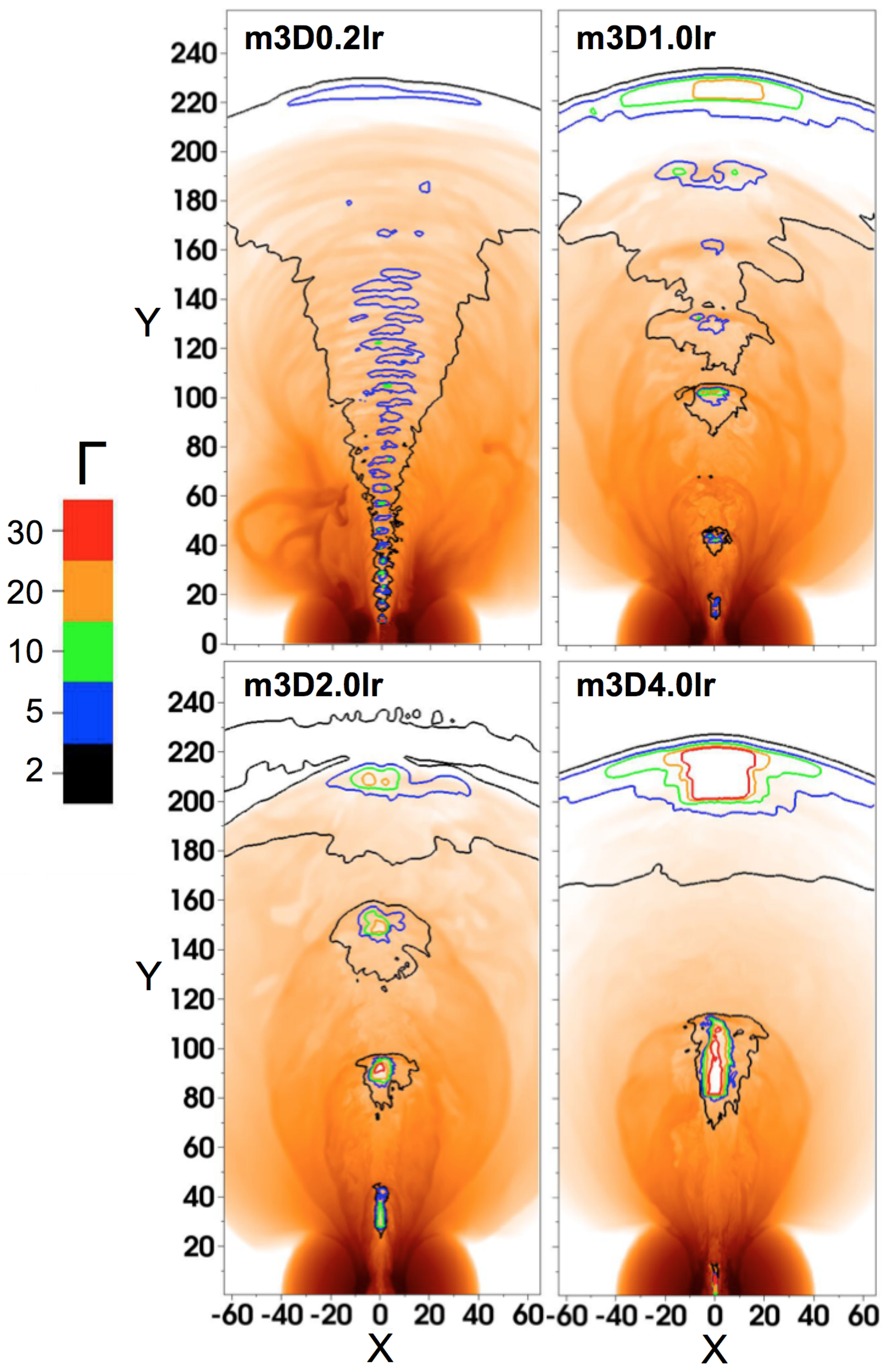}
  \caption{Lorentz factor contours ($\Gamma$=2, 5, 10, 20, 30) of the XY plane overlapped with the density stratification maps (same as in Figure~\ref{fig:fig1}) for models m3D0.2lr (at t=17.40s), m3D1.0lr (at t=13.93s), m3D2.0lr (at t=13.93s) and m3D4.0lr (at t=12.73s), (upper-left, upper-right, lower-left, and lower-right respectively). (A color version of this figure is available in the online journal.)} \label{fig:fig3}
\end{figure}

The Lorentz factors ($\Gamma$) of the active and quiescent times of the 3D models are shown in Figure~\ref{fig:fig3} (we must note that the timeframes shown were hand-picked in order to have $\approx$ the same cocoon size and to be able to compare them). Clearly the variability in the central engine generates high- and low-$\Gamma$ regions. In accordance with \citet{lc14}, the obtained $\Gamma$ distribution from the 3D models shows that the active periods produce low-density and high-$\Gamma$ regions contrary to the quiescent periods that have higher density and lower $\Gamma$ regions. The $T=4$~s model has two active regions (which reach at least $\Gamma$=30) and are separated by a large quiescent region which has Lorentz factors of order 2. The $T=2$~s model produces 4 pulses with high-$\Gamma$ ($\sim$20) separated by low-$\Gamma$ quiescent epochs. The one second period model has seven high-$\Gamma$ regions and the 0.2s period models has more than twenty high-$\Gamma$ regions. The active regions from the latter two models reach $\Gamma \ge$20, and 10 (respectively). The control model, model with the central engine active at all times, produced a single low-density high-$\Gamma$ region which reached values very similar to the $T=4$~s ($\Gamma \ge$30), thus there is a correlation between the period of pulses and their average $\Gamma$ values. 

\begin{figure}
  \centering
  \includegraphics[width=0.6\linewidth]{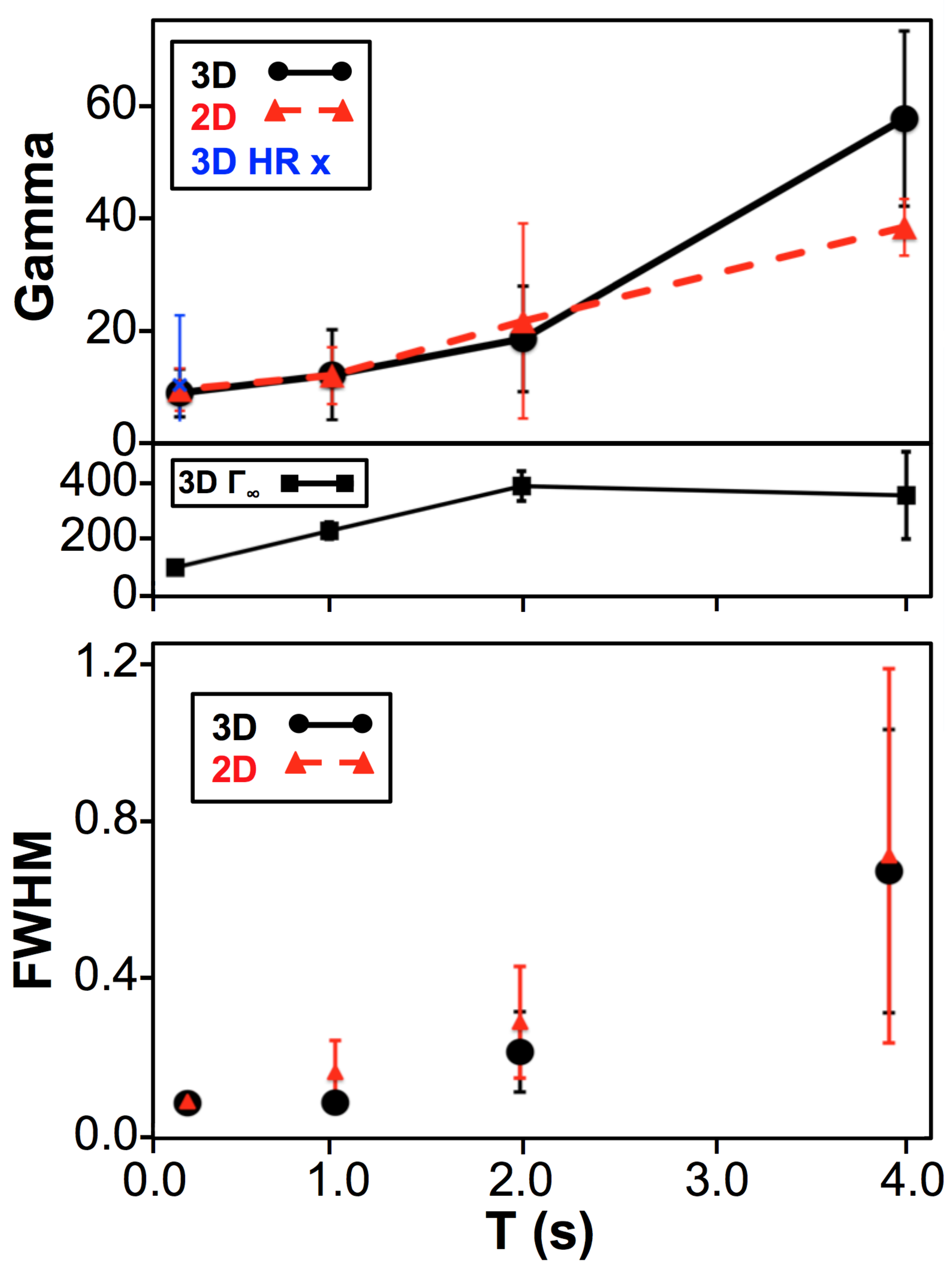}
  \caption{Upper panel: Pulses average Lorentz factor ($\pm$1$\sigma$
    error bars) of the 3D and the 2D episodic models (black solid line
    with black circles and red dashed line with red triangles,
    respectively). The pulses average Lorentz factor of the 3D HR
    model is also included (blue cross). Middle panel: Pulses maximum potentially 
    achievable Lorentz factor ($\pm$1$\sigma$ error bars) of the 3D episodic models.
    Bottom panel: Pulses average FWHM ($\pm$1$\sigma$ error bars) of the 3D and 2D 
    episodic models. (A color version of this figure is available in the online
    journal.)}
  \label{fig:fig4}
\end{figure}

To better understand the correlation between the pulses $T$ and $\Gamma$, in the upper panel of Figure~\ref{fig:fig4}, we plot the pulses average $\Gamma$ value for each of the 3D models (including its $\pm$1$\sigma$ error bars) for when the jet has reached 2$\times$10$^{11}$~cm. The best fit line for such correlation (with R$^2$=0.92) is $\Gamma$=13.01Hz$\times T$+0.24. In the middle panel of Figure~\ref{fig:fig4} we plot the pulses maximum potentially achievable Lorentz factor $\Gamma_\infty$ ($\Gamma_\infty=\Gamma[1+p/(\rho\,c^2)]$). Pulses from the long period engines reach their maximum potential acceleration ($\Gamma_\infty=400$), while pulses from central engines with faster variability do not. The latter is due to baryon loading as the jet drills through the progenitor. Faster variability increases the baryon load loading and thus the $\Gamma_\infty$ for the shortest variability period is significantly reduced ($\Gamma_\infty \sim 100$). Finally, there is a correlation between the average full width at half maximum (FWHM) of the pulses of the 3D episodic models and the correspondent period (FWHM$\propto T$), as shown in the bottom panel of Figure~\ref{fig:fig4}, this will be further discussed in Section~\ref{sec:HRLC}.

\subsection{Resolution effects and light curves}\label{sec:HRLC}
\begin{figure}
  \centering
  \includegraphics[width=0.85\linewidth]{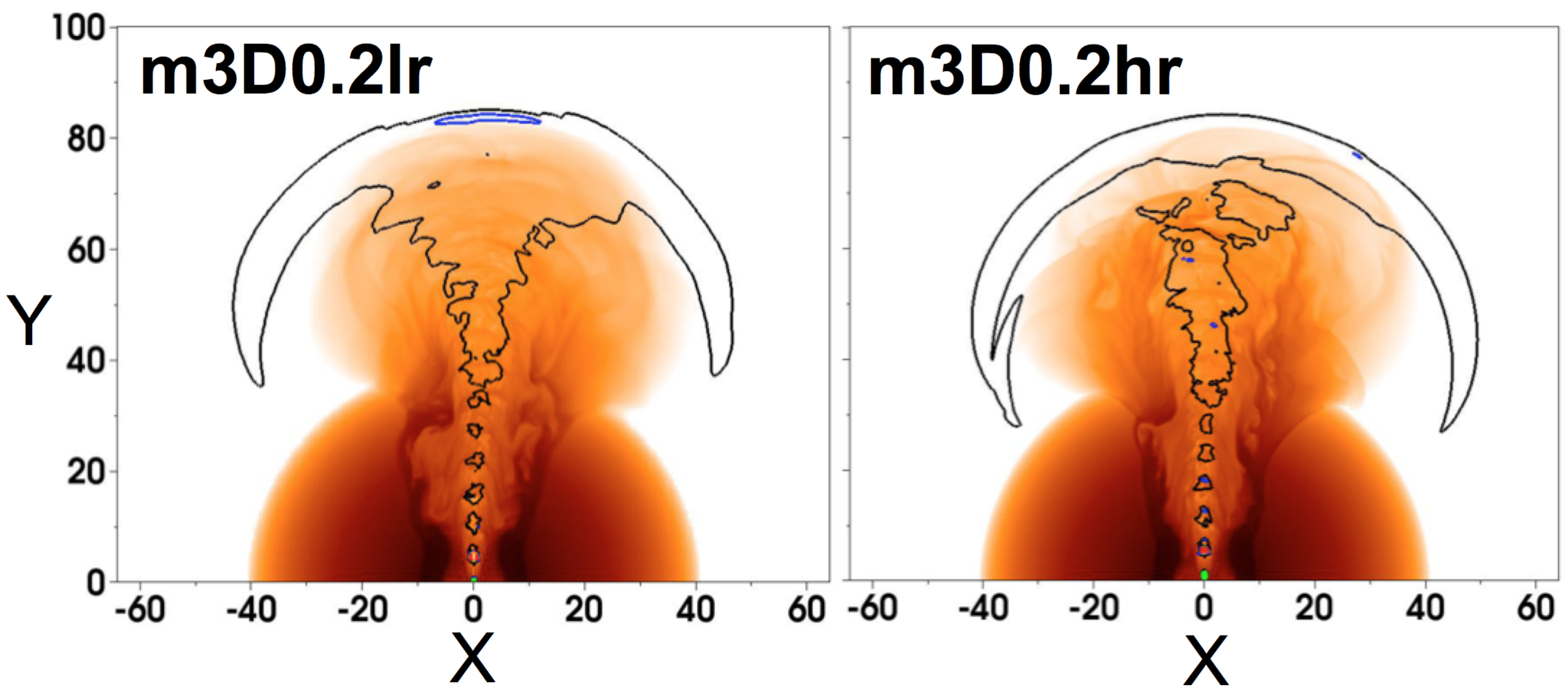}
  \caption{Lorentz factor contours overlapped with the density
    stratification maps of the XY plane (same as in Figure~\ref{fig:fig3}) for models
    m3D0.2lr (at t=12.47s) and m3D0.2hr (at t=12.63s) (left and right
    panels respectively).}
  \label{fig:fig5}
\end{figure}

To check for resolution effects and to assess the reliability of our results, we performed an extra three dimensional 0.2~s period model with a broader high-resolution region (m3D0.2hr). The HR model had the same input and boundary conditions as the LR model but contrary to the LR model the cocoon in the HR was solved entirely with its finest grid level (4$\Delta$). The t$_{\rm{bo}}$ for the HR model (t$_{\rm{bo}}$=11.20s) is practically the same as that from the LR model (less than 2\% difference, see Table 1). In Figure~\ref{fig:fig5}, we show the comparison of the density map (with $\Gamma$ contours) between the HR and LR for the $T=$0.2s model once both models have broken out of the star. Akin to \citet{lc13} the HR model not only presents more turbulence but also evolves slightly slower ($\sim$1\% slower) because of a larger amount of turbulence within the cocoon enhanced by the finer resolution. Due to numerical limitations the HR was only followed up to 12.8 seconds. By this time, the HR model has a higher $\Gamma$ value at the base of the jet ($\Gamma$=30) and lower value at the head of the jet ($\Gamma$=2) compared to the LR case ($\Gamma$=10 and 5, respectively).

Independently of the turbulence and $\Gamma$ value differences, both HR and LR models not only have very similar t$_{\rm{bo}}$ but, as can be seen in Figure~\ref{fig:fig5}, also show the same large scale structures. The evolution of the variable jet is, on large scales, basically the same. The high resemblance between the HR and LR three-dimensional models is further viewed in the top panel of Figure~\ref{fig:fig4} where the average $\Gamma$ values of the pulses for both HR and LR models are within one sigma from each other. Hence, since this study is devoted into analyzing the grand scale properties of the 3D variable jet drilling through the progenitor and ISM we need not require such large fine resolution regions as the HR case and we can ignore the possible resolution effects in our LR simulations.

\begin{figure}
  \centering
  \includegraphics[width=0.75\linewidth]{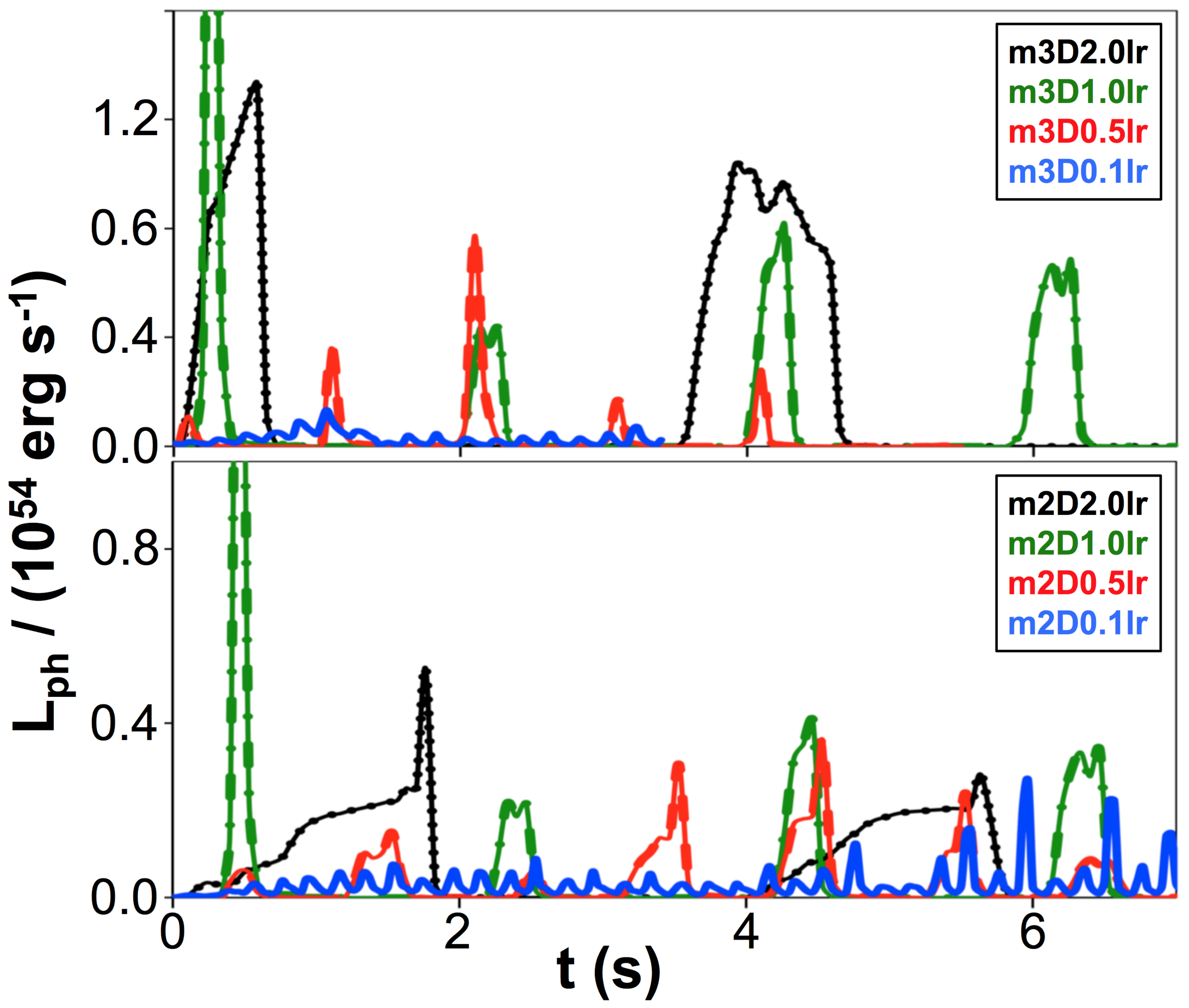}
  \caption{Light-power curves of the 3D (upper panel) and 2D (bottom
    panel) variable jet models (the time is normalized to the moment
    the jet reaches R$_{{\rm{ph}}}$, and also $\theta$ =
    1$^{{\rm{o}}}$). The $T=$4s, 2s, 1s and 0.2s models are shown in
    black dots, green dashes, red dot-dashed, and solid line all
    connected with a thin line with its respective color. (A color
    version of this figure is available in the online journal.)}
  \label{fig:fig6} 
\end{figure}

For each 3D episodic model, we estimated the light-power curves (LPCs) at r$_{{\rm{obs}}}$=2.2$\times$10$^{11}$cm and for an observer lying at 1$^o$ from the jet axis. The LPCs were calculated following the same procedure as in \citet{mor10}, see Figure~\ref{fig:fig6}. We must note how the time shown in the x-axis corresponds to the time elapsed after the first pulse reached r$_{{\rm{obs}}}$. The variable jet behavior is present in the LPCs and the active epochs are characterized by luminosities in the range L=10$^{53}$-10$^{54}$~erg~s$^{-1}$.

An important feature of the LPCs of Figure~\ref{fig:fig6} is that the pulse durations are significantly smaller than half the period of the injected variability. This is also noticeable in the lower panel of Figure~\ref{fig:fig4} where the average FWHM of the pulses for the 3D models are shown. For the three-dimensional $T=$4~s model for example, the average FWHM of the pulses obtained at r$_{{\rm{obs}}}$ is 0.660s (and an average of the quiescent epoch equal to 3.340s). The $T=$2~s model has an average FWHM value for the pulses equal to 0.209s (and and average FWHM for the quiescent epochs equal to 1.791s), 0.082s (and 0.918s) for the one second period model, and 0.08s (and 0.12s) for the 0.2s period model. This difference is due to the fact that the stellar material can re-fill the jet path in between pulses if the engine is turned off for a long time. As Figure~\ref{fig:fig3} shows, the inter-pulse space is filled by slow, dense material, and the jet material needs to work its way against it. Only for the faster engine the effect is not seen (see bottom panel of Figure~\ref{fig:fig4}).

\subsection{Two-dimensional models}\label{sec:2D}
The final intention of this study is to compare the results from the numerical simulations with real data from GRB observations. To do so, we require to construct a large synthetic dataset from the numerical models with comparable amount of pulses and quiescent periods as the observations. Unfortunately, constructing such synthetic dataset form the 3D simulations is out of the computational possibilities. Thus, we are left with using less expensive 2D episodic models. First though, we need to check whether the 2D models reproduce the main morphology of the 3D variable jet models, and then build the large synthetic dataset from these fast models to finally compare the results with the
observations.

In order to validate whether or not the 2D simulations reproduce the same morphological behavior as the 3D models we ran a first set of 2D simulations (four in total) with the same resolution, numerical domain, boundary conditions, periods as the 3D LR models, and were carried out in cylindrical coordinates. The first important point to mention concerning the 2D variable jet models is that they all managed to break out of the stellar progenitor. The break out times for the $T=$0.2s, 1s, 2s and 4s two-dimensional models were 5.60s, 10.07s, 11.93s, and 12.34s (respectively). These t$_{\rm{bo}}$ are once more consistent with previous two- and three- dimensional GRB jet studies (already mentioned before), as well as the analytic solution of \citet{brom11}. However, the results from these runs display a correlation between engine variability and break-out time that is opposite to the one found in 3D.

\begin{figure*}[ht]
  \centering
  \includegraphics[width=0.75\linewidth]{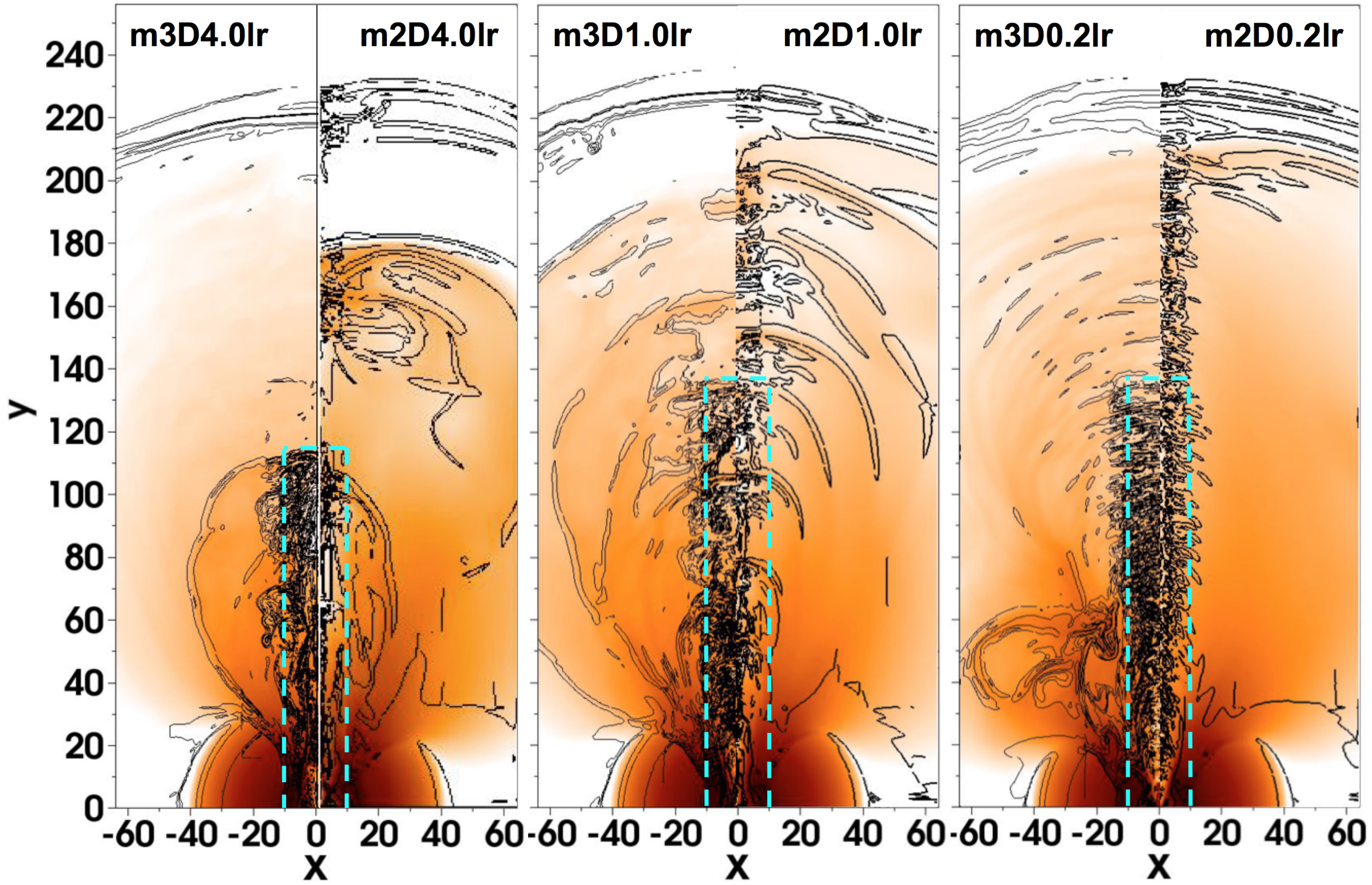}
  \caption{Schlieren maps for the $T=$4.0s, 1.0s, and 0.2s models using 3D (left panels in each figure) and 2D (right panels in each figure) simulations. (A color version of this figure is available in the
    online journal.)}
  \label{fig:fig7}
\end{figure*}

The origin of such a strikingly opposite behavior between the variability of the engine and its correspondent t$_{\rm{bo}}$ for 2D and 3D is not entirely clear, but is most likely related to the enhanced turbulence seen in 3D simulations versus 2D ones (for further details see Figure~\ref{fig:fig7}). For a continuous engine, the 3D simulation is faster due to the possibility of the jet head to skirt around the high-density area in front of the jet itself \citep{zwh04, lc13}. For an unsteady engine, instead, propagation in 3D is more difficult due to the increased amount of turbulence excited by the unsteady engine. In any of the cases, the difference seems to be related to the propagation of the head of the jet only. Once the channel has been drilled, 2D and 3D simulations behave remarkably similarly. As for the 3D models the episodic 2D models show a correlation between the average $\Gamma$ value and $T$: $\Gamma$=7.77Hz$\times T$+5.84 (with R$^2$=0.98) and between the average FWHM of the pulses and $T$ (see upper panel of Figure~\ref{fig:fig4}). Even though the slope of 2D correlation between $\Gamma$ and $T$ is smaller than its respective 3D case, it falls within the 1$\sigma$ error bars of the 3D values. The average FWHM of the pulses from 2D episodic models also falls within the 1$\sigma$ error bars of the 3D values (see the lower panel of Figure~\ref{fig:fig4}). The latter is also visible in the lower panel of Figure~\ref{fig:fig6} where the LPCs of the 2D models are shown. The details of the average FWHM of the pulses and its correspondent $\sigma$ error bars for both the 3D and 2D variable jet models are shown in Table 2.

\begin{table}
\centering
\tabletypesize{\footnotesize}
\caption{Model characteristics}
\begin{center}
\begin{tabular}{cccc}
\hline
Model & FWHM (s) & $\sigma$ (s)&\\
\hline
m3D0.2lr & 0.082 & 0.011 \\
m2D0.2lr & 0.085 & 0.023 \\
m3D1.0lr & 0.082 & 0.003 \\
m2D1.0lr & 0.158 & 0.080 \\
m3D2.0lr & 0.209 & 0.100 \\
m2D2.0lr & 0.283 & 0.140 \\
m3D4.0lr & 0.660 & 0.354 \\
m2D4.0lr & 0.699 & 0.467 \\
\hline
\end{tabular}
\end{center}
\label{default}
\end{table}

Finally, we analyzed the Schlieren maps ($\nabla$(log$_{10}(\rho)$) for both the 3D and 2D models in order to trace how the turbulence and shocks differs between the 2D and 3D cases. Figure~\ref{fig:fig7} shows the comparison between the Schlieren maps. Each panel shows the 3D XY plane on the left hand side while the right hand side shows the 2D case. From left to right, the three panels show the $T=$4s, 1.0s and 0.2s models. We must note that the timeframes for all models were handpicked so that the cocoon has expanded up to r$_{{\rm{obs}}}$. The 2D episodic models show larger bow shocks and less turbulence within the cocoon compared to the 3D cases. Never the less, as the variable 2D jet drills through the progenitor the turbulence and bow shocks from the $\approx$lower half of the jet resembles very much that from its respective 3D case (see the regions within the cyan dashed lines in Figure~\ref{fig:fig7}). Adding then the similarity between the break out times, the correlation between average $\Gamma$ and $T$, the correlation between the average FWHM and $T$, and the resemblance of the Schlieren maps within a small angle $\theta_{{\rm{obs}}}$ between the 3D and 2D models, we conclude that the 2D models do reproduce the large scale morphology of the 3D models. Thus, we build the extensive synthetic dataset from a large number of 2D simulations and compare it with the observations. We also notice how in both 2D and 3D models a bow shock is present ahead of each pulse, confirming that each jet pulse is working against the stellar material, not only the first pulse at the head of the jet.

\subsection{Synthetic data set and comparison with the observations}\label{sec:simsvsobs}
NP02 analyzed the temporal behavior of 68 long GRBs. The total of the pulses and quiescent times in these bright bursts was in total: 1330 pulses and 1262 quiescent intervals. Their main result was the detection of an asymmetry between the distribution of the pulse durations and the distribution of the duration of the quiescent periods. Our 2D and 3D simulation results show that pulse duration from periodic engines is systematically decreased, and the quiescent period duration consequently increased. Such pulse shortening (and quiescence duration increase) is due to the fact that the quiescent regions are refilled by stellar material (specially in long quiescent epochs where there the stellar material has more time to refill the funnel before the subsequent pulse crafts the funnel again). This point is backed by the presence of clear bow shocks in the Schlieren maps of the intermittent jets (shown in Figure~\ref{fig:fig7}) and could thus be the mechanism at the base of the asymmetry detected by NP02. In order to quantify the effect on a more realistic dataset of LPCs, we constructed a large synthetic dataset with comparable amount of GRB pulses and quiescent periods (with non-periodic engines). For this, we ran twenty 2D variable jet simulations, each one with random pulse and quiescent epoch durations (all with the same resolution and boundary conditions as the 2D models discussed in Section~\ref{sec:2D} (see Table 1 for further details). In each model the injected pulses and quiescent periods varied between 0s and 4s, resulting in at least five pulses and five quiescent times in each 2D random variable jet model. In total we injected 240 pulses and 246 quiescent epochs in the 20 simulations performed.

\begin{figure}
  \centering
  \includegraphics[width=0.8\linewidth]{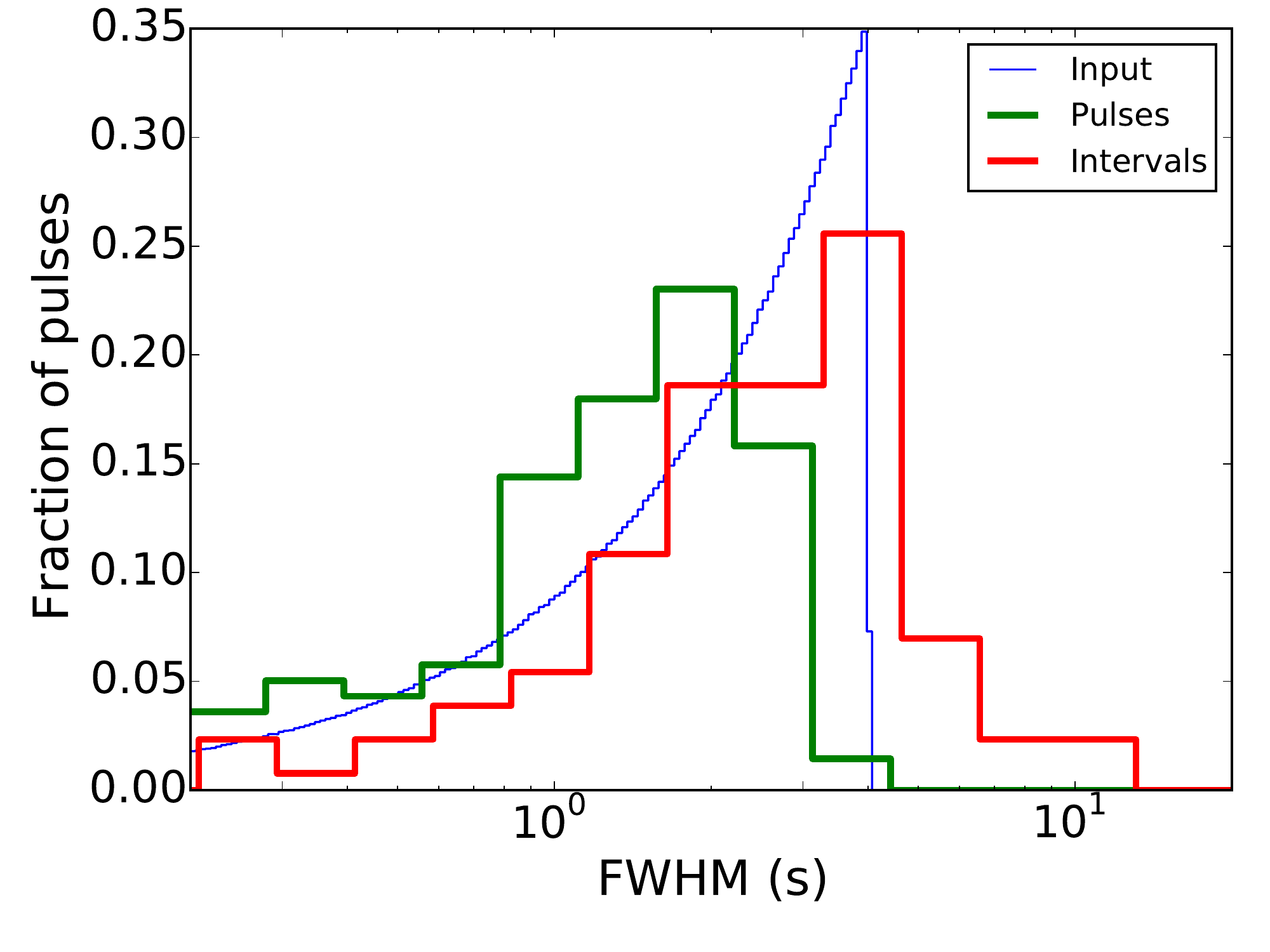}
  \caption{Histogram of the measured pulse (green) and quiescent intervals (green) duration distributions. The average input distribution for both the pulses and quiescent intervals is also shown (blue).}
  \label{fig:fig8}
\end{figure}

The normalized histogram of the measured duration distribution of both the pulses and quiescent epochs is shown in Figure~\ref{fig:fig8}. As for the 3D and 2D models with equal active and quiescent durations, the synthetic dataset obtained from the 2D models with random pulse and quiescent durations also shows a decrease (and increase) in the pulses (and quiescent) distributions obtained at r$_{{\rm{obs}}}$. Figure~\ref{fig:fig8} shows a clear asymmetry in the obtained distributions, with the quiescent interval durations extending to much longer times than the pulses. This replicates the NP02 result. However, it was not possible for us to completely reproduce their analysis, since their definition of pulse duration was based on the background level and our LPCs do not have background. We therefore computed the pulse duration as its FWHM and the quiescent period duration as the time during which the emission was less than 10\% of the peak of the preceding pulse. NP02, instead, computed the pulse duration as the interval between two points (on each side of the peak) that are higher than the background by 1/4 of the peaks height or by the minimum between two neighboring peaks (if the latter is higher), and the duration of quiescent intervals as the length of time during which no detectable flux is present. Due to the different methodology (in how the pulses and quiescent epochs are computed) between NP02 and this study, a more quantitative comparison is not possible.

\begin{figure}
  \centering
  \includegraphics[width=0.8\linewidth]{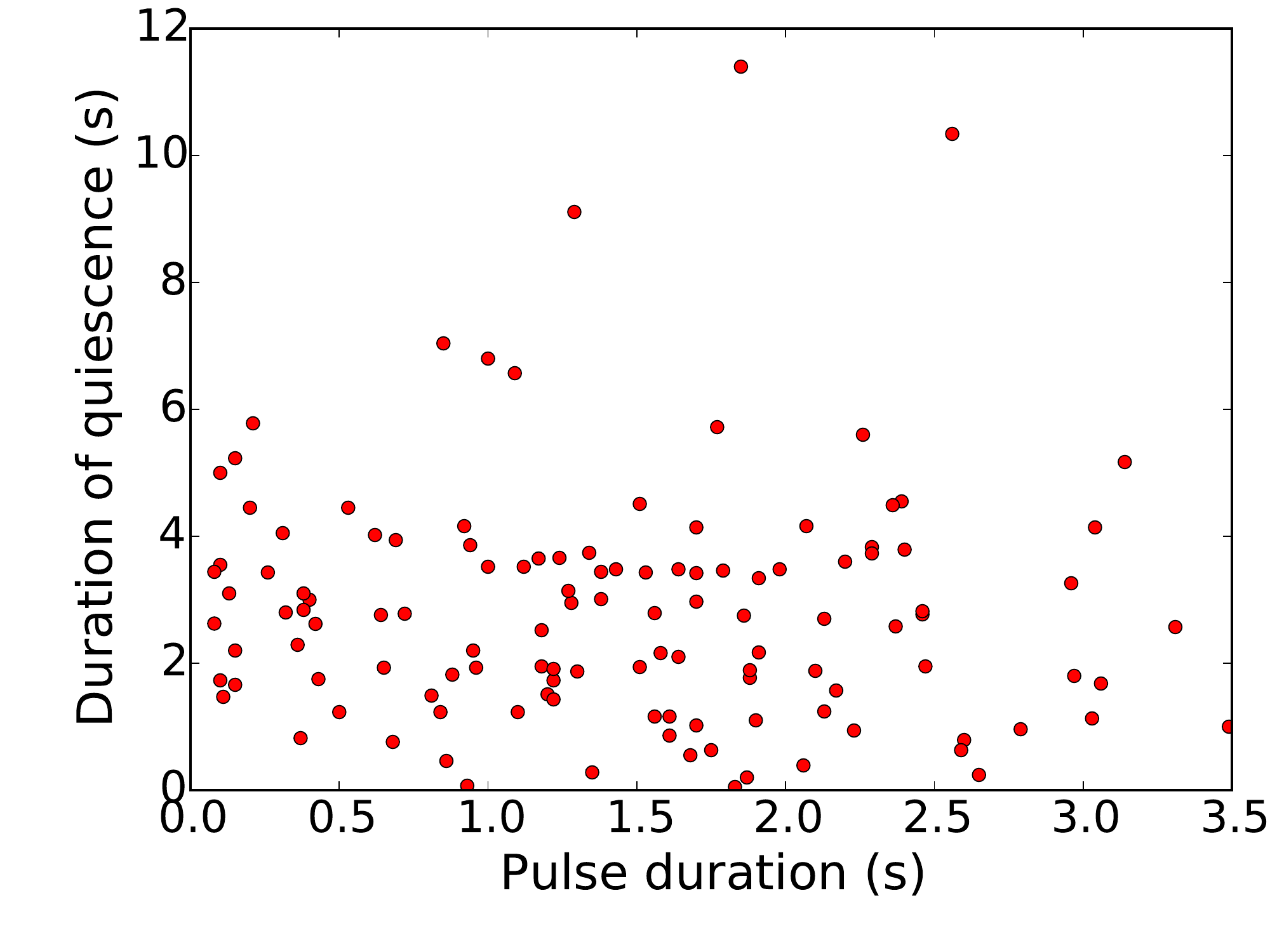}
  \caption{Scatter diagram of the duration of quiescent periods versus the duration of the following pulse.}
  \label{fig:fig9}
\end{figure}

NP02, in their dataset, also found a positive correlation between the duration of the quiescent periods and the subsequent pulse. We thus searched for a possible correlation between the quiescent intervals and the following pulse in our synthetic dataset. Figure~\ref{fig:fig9} shows a scatter plot of the preceding quiescent interval duration versus the pulse duration. We find no statistically significant trend. Such lack of correlation between the quiescent periods and pulses could be due to the bow shocks and the reduced amount of turbulence that are present in the 2D models. 

\section{Summary and Conclusions}\label{sec:conc}
We presented a set of 2D and 3D numerical simulations of the propagation of relativistic GRB jets in progenitor stars and beyond. These new simulations confirm the previous 2D results that episodic jets manage to drill and break out of the progenitor and that the active periods are low-density and high-Gamma (contrary to the quiescent periods which have higher densities and lower Lorentz values).

We find that there are correlations between the period of the engine variability and the average Lorentz factor of the pulses and also between the period of the engine variability and the time the jet takes to break off the stellar surface. We show how both the average Lorentz factor of the pulses and the correspondent Lorentz factor at complete acceleration are affected by the engine variability. The faster the engine variability the more loaded with baryon is the jet. Still, a more detailed treatment of the jet injection is required to better understand how the baryon loading affects the light curve. A model in which the engine luminosity is fairly constant but some other property varies (e.g., magnetization or velocity stratification) would provide a more successful outcome by keeping the funnel open while creating variability as a result of more efficient dissipation. We also found that the propagation through the stellar progenitor has the effect of shortening the pulse durations and expanding the quiescent times, especially for engines with long variability period. This behavior seems to be due to the fact that the jet channel refills with stellar material if the engine is turned off for too long a time (more than a fraction of a second). A comparison of pulse properties with bright BATSE GRBs (NP02), however, is not entirely successful since our synthetic light curves do not reproduce the correlation between the duration of pulses and of the quiescent time preceding the pulses. Such correlation is probably due to the central engine itself and not a consequence of the propagation of the jet through the progenitor star.

\textbf{Acknowledgements}
We thank S.E. Woosley and A. Heger for making their pre-SN models available. The software used in this work was in part developed by the DOE-supported ASC/Alliance Center for Astrophysical Thermonuclear Flashes at the University of Chicago. This work was supported in part by the CONACYT Research Fellowship (DLC), and the NSF grant AST1333514 (BJM).


\end{document}